\documentclass[prc,preprint,superscriptaddress]{revtex4}
\usepackage{graphicx}
\usepackage{amsmath,bm}
\usepackage[T1]{fontenc}
\usepackage[english]{babel}
\UseRawInputEncoding
\begin{document}

\title{{  \large Nonlinear corrections to the nuclear heavy flavor structure functions }}
\author{{\bf F.Abdi\footnote{E-mail: fariba.abdi @ razi.ac.ir}}}

\author{{\bf B.Rezaei\footnote{E-mail: brezaei@ razi.ac.ir}}}

\affiliation{  Department of Physics, Razi University, Kermanshah 67149, Iran}
\begin{abstract}
We  study numerically the small-$x$ behaviour of  the nuclear gluon distribution function $ G^A(x,Q^2)$ at next-to-leading order (NLO)  approximation
of the Gribov-Levin-Ryskin-Mueller-Qiu, Zhu-Ruan-Shen (GLR-MQ-ZRS)  nonlinear equation and quantify the impact of gluon recombination in the kinematic range
of $x \le 10^{-2}$ and $Q^2 \ge 5 \text{GeV}^2$ respectively.  The results are comparable to the Rausch-Guzey-Klasen (RGK) [J.Rausch, V.Guzey and M.Klasen, Phys. Rev. D 107, 054003 (2023)] nuclear gluon distributions and the nCTEQ15 parametrization at the corresponding   $Q^2$ values. Using the solutions of  $ G^A(x,Q^2)$ in the framework of the nonlinear GLR-MQ-ZRS evolution equation, the linear and nonlinear behavior of the charm  structure functions of nuclei per nucleon $F^{c\bar{c}-A}_2(x,Q^2)$ and  $F^{c\bar{c}-A}_L(x,Q^2)$ are considered. The results reveal that nonlinear corrections play an important role in charm nuclear reduced cross sections at small-$ x$ and low $Q^2$ values. The computed results are compared with experimental data from the H1 and ZEUS Collaborations.
\end{abstract}

\maketitle
\section{INTRODUCTION}

The process of  Deep-inelastic scattering (DIS) of high-energy leptons off hadrons, for studying the structure of hadrons in terms of their quark and gluon degrees of freedom, is  a key process \cite{DIS}. For both nucleons  and nuclei, available data from electron and neutrino DIS play an important role in modern determinations of parton distribution functions (PDFs), which describe the internal structure of hadrons \cite{CTEQ,Bailey,Hirai, Eskola,nCTEQ15,Khanpour}. In the future electron-ion collider (EIC), this process will be important again where the DIS of various nucleons and nuclear targets will be studied with high precision over a wide kinematic range \cite{EIC}. Further,  when searching for the transition between linear and nonlinear scale evolution of the parton density exact knowledge of nPDFs will be important. The nonlinear regime, known as saturation \cite{13,Marquet}, occurs at small-$ x$ and low  $Q^2$ values where the recombination of small-$x$ gluons becomes increasingly important. In lepton-nucleus ($\ell+A$) scattering such nonlinearities are predicted to be more pronounced than in lepton-proton ($\ell+p$) interactions \cite{17}. Establishing nonlinear effects can either be done by comparing the behavior of nPDFs extracted in different $x- Q^2$ regions with and without sizeable nonlinear effects \cite{14}. This is one of the key physics goals of an Electron-Ion Collider (EIC) \cite{EIC}.\\
The study of DIS of leptons on nuclei shows the appearance of a significant nuclear effect, which rejects the  simple idea of the nucleus as a system of quasi-free nucleons \cite{Arneodo1,Arneodo2,Rith}. This effect was first discovered by the European Muon Collaboration in the domain of valence quark dominance, so it was named as the EMC effect \cite{EMC}.
In perturbative quantum chromodynamics (pQCD), the two well-known parton evolution equations Dokshitzer-Gribov-Levin-Altarelli-Parisi (DGLAP) \cite{Dokshitzer,Gribov,Altarelli} and Balitsky-Fadin-Kuraev-Lipatov (BFKL) \cite{Fadin,Lipatov} serve as the essential tools for prediction of PDFs on their respective kinematic domains.\\
The gluon density becomes  dense in extremely small-$x$ at fixed $Q^2$. As a result, the gluons start to overlap in the transverse space. The correlative interactions between gluons become important in this region. In other words, the probability of recombining two gluons into one increases $ (g + g \rightarrow g)$, which will tame the increase of gluons and eventually lead to the formation of gluon saturation \cite{Blaizot,Marquet}.  This process is the inverse process of gluon splitting $(g \rightarrow g + g)$. The nonlinear evolution equations, including the Jalilian-Marian-Iancu-McLerran-Weigert-Leonidov-Kovner (JIMWLK) equation \cite{Jalilian1,Jalilian2,Iancu}, its mean-field approximation the Balitsky-Kovchegov (BK) equation \cite{BK1,BK2}, the Gribov-Levin-Ryskin-Mueller-Qiu (GLR-MQ) equation\cite{GLR1,GLR2}, and the Modified-DGLAP equation\cite{M-DGLAP}, are widely employed to investigate the saturation phenomenon. 
The GLR-MQ equation incorporates gluon recombination effects, calculated using the AGK cutting rules, into its formulation. This new evolution equation can serve as a tool to restore the Froissart bound as well as  the unitarity. The fundamental difference between nonlinear equation(GLR-MQ) and the linear DGLAP equation is due to the presence of a shadowing term in the GLR-MQ. The shadowing term in gluon density is coming from gluon recombinations inside the hadrons. There are several issues that seem to appear in the GLR-MQ equation:\\
(a) the application of the AGK cutting rule in the GLR-MQ corrections breaks the evolution kernels\cite{Zhu1};\\
(b) the nonlinear term in the GLR-MQ equation violate the momentum conservation \cite{Zhu2};\\
(c) the Double Leading Logarithmic approximation(DLLA) is valid only at small-$ x$ and the GLR-MQ corrections cannot smoothly connect with the DGLAP equation\cite{Zhu3}.\\
Zhu and his colleagues derived a new QCD evolution equation including parton recombination in the forward logarithmic ($Q^2$) approximation  using time-ordered perturbation theory (TOPT) instead of the AGK cutting rules. This new evolution equation is known as the GLR-MQ-ZRS equation \cite{Zhu1,Zhu2,Zhu3,Zhu4,Zhu5,Zhu6}. The violation of the momentum conservation of the GLR-MQ equations was resolved by the calculations of Zhu and Ruan \cite{Zhu2,Zhu4}. The effect of the nonlinear corrections is to moderate the $Q^2$ growth of PDFs at small-$ x$, while somewhat increasing the speed of evolution at moderate values of $x$ such that the total momentum is conserved. This discussion requires talking about the terms  dynamically generated shadowing and antishadowing terms which are  more familiar in the context of nuclear PDFs \cite{Klasen}.\\
To discover this  increasement is one of the basis of the envisioned physics programs of the EIC \cite{EIC,eic} and Large Hadron Electron Collider (LHeC) \cite{LHeC1,LHeC2} and a strong motivation to run the colliders with both light and heavy nuclei. The Future Circular Collider (FCC) \cite{FCC}, which would attain kinematics even further beyond those of the LHeC in electron-ion collisions, would naturally be able to put even stronger constraints on the nature and strength of nonlinear corrections to the evolution of PDFs.\\
The data \cite{data1,data2,data3} show  the nuclear structure function $ F^A_2 (x, Q^2)/A$ in nuclei are different from  the free nucleon structure function $F^N_2 (x,Q^2)$. This observation has prompted comprehensive global analyses of nuclear parton distribution functions (nPDFs). The first such fit was EKS98 \cite{EKS98}, also including Drell-Yan (DY) dilepton data, followed by HKM providing the first error analysis \cite{HKM}. Both studies were conducted at leading-order (LO) perturbative QCD. The first next-to-leading order (NLO) analysis was provided by nDS \cite{NDS}. The first to include RHIC dAu hadron-production data was  the  EPS08 analysis \cite{EPS08}. Nuclear effects in the nPDF and in the structure function $ F^A_ 2$  are expressed in the following  analytical  forms\cite{Arneodo1,RF2}
\begin{align}
	R^A_i &=\frac{xf^A_i(x, Q^2)}{A xf^p_i(x, Q^2)}\notag\\
	R^A_{F_2} &=\frac{xF^A_2(x, Q^2)}{A xF^p_2(x, Q^2)},
\end{align}
where, $A$ is the nuclear mass number. The modifications depend on the parton momentum fraction, which can be classified into four distinct kinematic regions. The Fermi motion region is dfeined for momentum fractions $x\gtrsim 0.8$ , the EMC region for $ 0.3 \lesssim x \lesssim 0.8$ , the antishadowing region for $0.1\lesssim x \lesssim 0.3$,  and  the shadowing region for $x\lesssim 0.1$(  NMC\cite{NMC}).\\
High-energy experiments at CERN and Fermi lab. have particularly investigated the small-$x$ regime ($x<0.1$), and show a systematic reduction  of the per-nucleon nuclear structure function $F_2^A(x,Q^2)/A$ relative to the proton structure function $F_2^p(x,Q^2)$ \cite{Paakkinen}. This phenomenon is known as nuclear shadowing effect and is associated with the modification of the target parton distributions so that
$xq^A(x, Q^2) < Axq^N(x, Q^2)$. The shadowing of gluons is also determined by \cite{Rg}
\begin{equation}
	R^A_g=\frac{xg^A(x, Q^2)}{A xg^p(x, Q^2)} .
\end{equation}
In this paper,  we examine the nonlinear corrections to the gluon density for light and heavy nuclei. We then explore the behavior of the charm structure functions across a wide
range of light and heavy nuclei at the EIC center-of -mass (COM) energy.\\

\section{Nonlinear corrections}

The primary distinction between the GLR-MQ and GLR-MQ-ZRS equations lies in the inclusion of specific terms. In the GLR-MQ equation, only the shadowing term is considered, whereas the GLR-MQ-ZRS equation incorporates both shadowing and antishadowing terms. Furthermore, the ZRS corrections to the Altarelli-Parisi equation, when combined with recombination functions within the Double Leading Logarithmic Approximation (DLLA), are detailed in Refs. \cite{Zhu3,Zhu1}.
We can be expressed GLR-MQ-ZRS in  a more appropriate manner using $ xg(x,\mu^2) \rightarrow  xg^A(x,\mu^2)=G^A(x,\mu^2)$ which represents the nonlinear corrections to the evolution of the nuclear gluon distribution,
\begin{align}\label{zrs1}
	\frac{dG^A(x,\mu^2)}{d\ln \mu^2} &= \left.\frac{dG^A(x,\mu^2)}{d\ln \mu^2}\right|_{DGLAP} 
	+ \frac{9}{2\pi}\frac{N^2_c}{N^2_c-1}f(\mu^2)\int_{\frac{x}{2}}^{\frac{1}{2}}\frac{dz}{z}\left(G^A(z,\mu^2)\right)^2 \notag \\
	&\quad - \frac{9}{\pi}\frac{N^2_c}{N^2_c-1}f(\mu^2)\int_{x}^{\frac{1}{2}}\frac{dz}{z}\left(G^A(z,\mu^2)\right)^2,
\end{align}
where $N_c$ is the number of color charges, $N_c = 3$, and $f(\mu ^2)= \alpha_s^2(\mu^2)/(R^2_A \mu^2)$. Here, $R_A = R\times A^{1/3} $ represents the nuclear size for a nuclear target with the mass number $A$.  The magnitude of the nonlinear effects depends on the value of the correlation length $R$.  This parameter controls the strength of the nonlinear effects and which can be associated with the size of the area in the transverse plane, where the partonic overlap leading to the nonlinear corrections becomes important. The quantity $1/(RQ)^2$ can be interpreted as the probability of overlap between two partons, referred to as the recombination scale. Here, $ 1/Q$ is the scale of a parton at momentum transfer $Q^2$ and $R$ is the maximum correlation length of two partons\cite{Zhu1,Guzey}. If the gluons are populated across the proton  then $R\approx 5\, \text{GeV}^ {-1}$. On  the contrary, if the gluons have hotspot like structure then $R\approx 2 \,\text{GeV}^{-1} $\cite{R25}. For a
nuclear target, From muonic X-ray spectroscopy, the finite-size effect of the nucleus can
be directly observed in the atomic energy levels, allowing the determination of the nuclear
charge radius. Experimental data for different nuclei indicate that the effective radius of a
nucleon inside the nucleus is about  $R = 1.25\, \text{fm}\approx 6.34 \,\text{GeV}^{-1}$, which is in fact more closely related to the
average nuclear density than to the actual size of the proton or neutron \cite{Krane}.\\
The first term on the right hand side of the Eq.(\ref{zrs1})  is the linear DGLAP equation at Double Logarithmic
Approximation (DLA). The second and third terms correspond to the antishadowing and shadowing effects, respectively, which arise from the correlated gluons within the hadronic structure. In  GLR-MQ-ZRS equation,  the strong growth of gluons generated by the linear term was tamed down by the shadowing and antishadowing terms. In  Eq.(\ref{zrs1}), the positive antishadowing effects  are separated from the negative shadowing effects because of the fact that they have different kinematical domains in $x$. Specifically, both the shadowing and antishadowing coexist in the region $ x\le 0.1$ whereas  only the antishadowing effects are present in the region $x>0.1$.\\
The nonlinear corrections to the nuclear gluon distribution function is derived by solving the GLR-MQ-ZRS equation, and its expression is  obtained as follows:
\begin{align}\label{GLR-MQ-ZRS2}
	G^A_{NL}(x,\mu^2) &= G^A_{NL}(x,\mu_0^2) + G_L^{A}(x,\mu^2) - G_L^{A}(x,\mu_0^2) \notag\\
	&\quad + \frac{9}{2\pi}\frac{N_c^2}{N_c^2-1} \int_{\mu_0^2}^{\mu^2} d\ln \mu^2\, f(\mu^2) 
	\int_{x/2}^{1/2} \frac{dz}{z} \left( G_L^{A}(z,\mu^2) \right)^2 \notag\\
	&\quad - \frac{9}{\pi}\frac{N_c^2}{N_c^2-1} \int_{\mu_0^2}^{\mu^2} d\ln \mu^2\, f(\mu^2)
	\int_{x}^{1/2} \frac{dz}{z} \left( G_L^{A}(z,\mu^2) \right)^2,
\end{align}
here $G_L^A(x,\mu ^2)$ and $G_{NL}^{A}(x,\mu_0 ^2)$ are the linear gluon distribution functions. The function $G_L^{A}(x,\mu^2)$  is defined in Ref. \cite{Gp} with the  substitution  $Q^{2} _s\rightarrow Q^{2}_{s,A} $, $\sigma_0 \rightarrow \sigma^A_0$, and is expressed in the following form:
\begin{equation}\label{GA}
	G_L^{A}(x,Q^2) = \frac{3\sigma^A_0}{4\pi^2 \alpha_s} \Big[-Q^2 e^{(-Q^2/Q^{2}_{s,A})} + Q^2_{s,A}(1-e^{(Q^2/Q^{A}_{s,A})})\Big]
\end{equation}
here $\sigma^A_0$ is defined to be $\sigma^A_0=A^{\frac{2}{3}}\; \sigma_0$  and
$Q^{2}_{s,A}$ is defined to be $Q^{2}_{s,A}=A^\frac{1}{3}\;Q_s$ \cite{sigma,Braz,Qs}. Here, $Q_s$ represents the transverse momentum scale, which exhibits a dependence on the variable $x$ \cite{Q2s}, expressed as:
\begin{equation}
	Q^2_s = 1\thinspace \text{GeV}^2 .\   (\frac{\bar{x}}{x_0})^{-\lambda}.
\end{equation}
The values of the parameters $\sigma_0 = 27.32$~mb, $\lambda = 0.248$, and $x_0 = 0.42 \times 10^4$ for the 4-flavor case were taken directly from Ref.~\cite{cte}. The term $\bar{x}$ used in the equations refers to the  modified Bjorken variable, which is defined as $ \bar{x} = x(1 + 4\,m_c^2/\mu^2)$.\\
Nonlinear corrections to the gluon distributions for $x<x_0 \ (=10^{-2})$, at the initial scale $Q_0^2\equiv \mu^2_0$, are obtained from the results in Ref. \cite{g-sat} as
\begin{equation}\label{sat1}
	G_{NL}^{A}(x,\mu^2_0)=G_L^{A}(x,\mu_0^2)\bigg\{1+\theta(x_0-x)\Big[G_L^{A}(x,\mu_0^2)-G_L^{A}(x_0,\mu_0^2)\Big]/G^{A}_{sat}(x,\mu_0^2) \bigg\}^{-1}, 
\end{equation}
in the leading shadowing approximation,  $G^{A}_{sat}(x,\mu_0^2)={16}/ {27\pi f(\mu_0 ^2) }$ where  $ f(\mu_0 ^2) =R^2\mu_0 ^2/\alpha_{s}(\mu_0 ^2)$ represents the value of the gluon distribution that would saturate the unitarity limit. We note that for $x \geq x_0 \,(=10^{-2})$, the linear and nonlinear gluon distributions exhibit the same behavior. Therefore, the form of Eq.~\ref{sat1}, which incorporates nonlinear corrections only for $x < x_0$, is well justified, as discussed in Ref.~\cite{g-sat}. \\

\section{heavy quark structure functions at
	small $X$ in nuclei}

In the small x region, the heavy quark structure functions in the collinear generalized  double  asymptotic scaling (DAS) approach presented in Ref. \cite{DAS} can be generalized to nuclei, in the following form
\begin{equation}
	F_k^{\textit{Q}\bar{\textit{Q}}-A} (x,\mu^2) = e_\textit{Q}^2 \sum_{n=0} (\frac{\alpha_s}{4\pi})^{n+1}C^{(n)}_{k,g}(x,\xi)\otimes G^A(x,\mu^2), (k=2,L)
\end{equation}
where $n$ denotes the order in running coupling $ \alpha_s$ are the Wilson coefficient functions and $\mu$ is the renormalization scale. Here, $ e_\textit{Q}^2$, is the squared charge of the heavy quark and $m_\textit{Q}$ is the heavy quark
mass. The symbol $\otimes$ is the Mellin convolution
\begin{equation}
	C^{(n)}_{k,g}(x,\xi)\otimes G^A(x,\mu^2)=	\int_{ax}^{1} dy\frac{x}{y^2}C^{(n)}_{k,g}(\frac{x}{y},\xi) G^A(y,\mu^2)
\end{equation}
where $a=1+4\xi$ , $\mu^2 = Q^2+4m^2_Q$ and $ \xi= m^2_\textit{Q}/Q^2$. Here, $C_{k,g}$ is the collinear Wilson coefficient function in the high energy regime\cite{Bkg}. The coefficient function at LO is defined as
\begin{align}
	&\notag C_{2,g}^{(0)}(z,\xi)=-2\beta\left[ 1-4z\left( 2-\xi\right) \left( 1-z\right) -\left( 1-2z\left( 1-2\xi\right) +2z^2\left( 1-6\xi-4\xi^2\right) \right) L(\beta)\right],\\ 
	&	C_{2,L}^{(0)}(z,\xi)=8z^2\beta\left[ \left( 1-z\right) -2z\xi L(\beta)\right],  
\end{align}
where $ \beta^2=1-4\xi z/(1-z), L(\beta)=\frac{1}{\beta}(\ln\frac{(1+\beta)}{1-\beta})	$ and $z=x/y$. The NLO coefficient functions are currently available as computer codes \cite{b2Lg}. Nonetheless,  within the high-energy limit ($\xi \ll1$) we can used the compact form of these coefficients according to the Ref. \cite{bkg} (see Appendix).\\
Thus, the heavy quark structure functions can be derived by accounting for nonlinear effects, in the following form
\begin{equation}
	F_{k,NL}^{\textit{Q}\bar{\textit{Q}}-A} (x,\mu^2) = e_\textit{Q}^2 \sum_{n=0} (\frac{\alpha_s}{4\pi})^{n+1}\int_{ax}^{1} dy\frac{x}{y^2}C^{(n)}_{k,g}(\frac{x}{y},\xi)\, G_{NL}^A(y,\mu^2).
\end{equation}
Therefore, the reduced cross section of heavy quarks is defined in terms of the nonlinear structure functions of heavy quarks in nuclei  is defined by the following form
\begin{align}
	\sigma^\mathnormal{{\textit{Q}\bar{\textit{Q}}-A}}_r(x, \mu^2)& \notag=\frac{d^2	\sigma^{\textit{Q}\bar{\textit{Q}}-A}}{dxd\mu^2}.\frac{xQ^4}{2\pi \alpha^2_sY_+}\\
	\notag &=F_2^{\textit{Q}\bar{\textit{Q}}-A}(x,\mu^2)-\frac{y^2}{Y_+}F_L^{\textit{Q}\bar{\textit{Q}}-A}(x,\mu^2)\\
	&=F_2^{\textit{Q}\bar{\textit{Q}}-A}(x,\mu^2)[ 1-\frac{y^2}{Y_+}R^{\textit{Q}\bar{\textit{Q}}-A}],
\end{align}
where
\begin{equation}
	R^{\textit{Q}\bar{\textit{Q}}-A}= \frac{F_L^{\textit{Q}\bar{\textit{Q}}-A}(x,\mu^2)}{F_2^{\textit{Q}\bar{\textit{Q}}-A}(x,\mu^2)}= \frac{C^{(n)}_{L,g}(x,\xi)\otimes G^A(x,\mu^2)}{C^{(n)}_{2,g}(x,\xi)\otimes G^A(x,\mu^2)}.
\end{equation}	
	
\section{Results and Conclusion}

\begin{figure}[htbp]
		\centering
		\includegraphics[width=16cm]{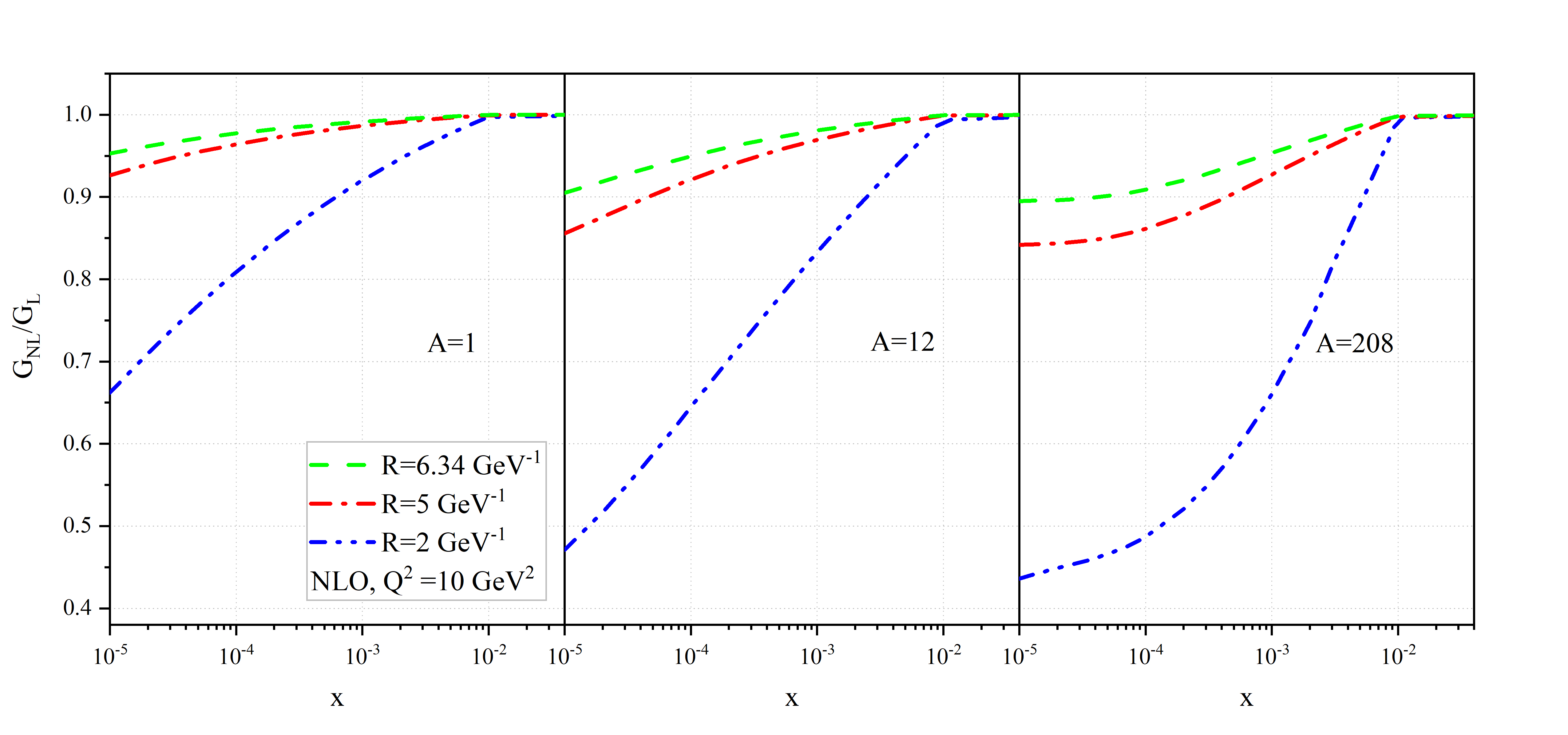}
		\caption{Ratios between the PDFs evolved using the nonlinear (Eq.\protect\ref{GLR-MQ-ZRS2}) and linear (Eq.\protect\ref{GA}) evolution equations at $Q_0 = 2~\text{GeV}$ are shown as functions of $x$ for different values of $R$. The left, middle, and right plots correspond to $A = 1$, $12$, and $208$, respectively.}
		\label{Fig1}
		\end{figure}

\begin{figure}[htbp]
	\centering
	\includegraphics[width=\textwidth]{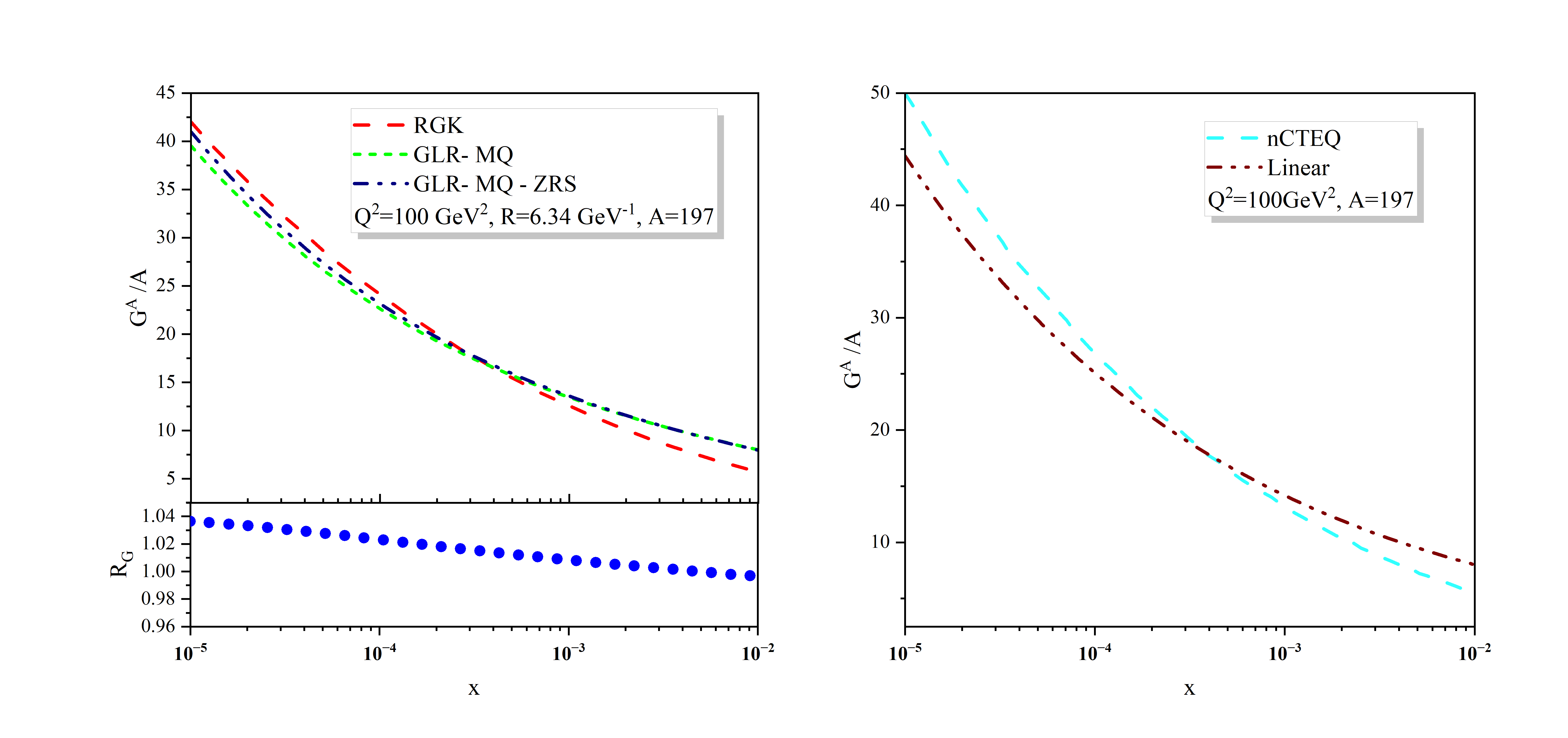}
	\caption{The linear and nonlinear gluon distribution function for the nucleus of Au-197. 
		Left panel: gluon distributions per nucleon and $R_{G}$ ratio ($R_G=\frac{G^A_{\rm GLR-MQ-ZRS}}{G^A_{\rm GLR-MQ}}$) as a function of $x$ for $Q^2 = 100\, \text{GeV}^2$. 
		Right panel: linear gluon distribution function as a function of $x$ at $Q^2 = 100\, \text{GeV}^2$. 
		Results are compared with the nuclear PDF set nCTEQ15 \cite{nCTEQ15} and the RGK method \cite{Guzey1} at the corresponding $Q^2$.}
	\label{Fig2}
\end{figure}

The running coupling in leading-order (LO) and next-to-leading-order (NLO) approximations is defined by the following expressions:
\begin{align}
	\alpha_s(\mu^2)&=\frac{4 \pi}{\beta_0\ln(\mu^2/\Lambda^2)}\qquad  \qquad \ \ \ \ \ \ \ \ \ \  \qquad\text{(LO)}\notag\\
	\alpha_s(\mu^2)&=\frac{4 \pi}{\beta_0\ln(\mu^2/\Lambda^2)} [1- \frac{\beta_1\ln\ln(\mu^2/\Lambda^2) }{\beta_0^2\ln(\mu^2/\Lambda^2)} \quad \text{(NLO)},
\end{align}
where $\beta_0= \frac{1}{3}(33-2N_f)$, $\beta_1= 102-\frac{38}{3}N_f$ are the one-
loop, two-loop  corrections to the QCD $\beta$-function. The variable $\Lambda$ the QCD cut- off parameter. The QCD parameter $\Lambda$ is extracted from the running coupling $\alpha_s(Q^2)$, where $\Lambda_{\text{QCD}}$ $0.12\ (0.292)\ \text{GeV}$
yields $\alpha_s(M_Z) = 0.118$ for the LO and NLO coupling, with the number of active flavors being $N_f = 4$ \cite{Tanabashi}. 
The mass of the charm quark is set to $m_c = 1.65\ \text{GeV}$ \cite{mc}.\\
In this paper, we have solved numerically the nonlinear GLR-MQ-ZRS equation in the kinematic range of $x<10^{-2}$ and $Q^2>5 \ \text{GeV}^2$ respectively. We exhibit our calculation results that have been attained for the linear and nonlinear gluon distribution function  $G^A(x, Q^2)$, reduced cross-section $\sigma_r^{c\bar{c}-A} (x,Q^2)$, the structure functions $F^{c\bar{c}-A}_2(x,Q^2)$ and $F^{c\bar{c}-A}_L(x,Q^2)$ and their ratios with respect to variables $x$ and $Q^2$ for nuclei at the LHeC and EIC COM energies at the LO and NLO approximations.\\
Fig.\ref{Fig1} quantifies the magnitude of the nonlinear corrections as a function of the mass number $A$. The figure displays the ratios of the nonlinear to linear gluon distribution functions, denoted by $G_{\text{NL}}(x,Q^2)/G_{\text{L}}(x,Q^2)$, as a function of $x$ for various nuclei including $\text{C}-12$, $\text{Pb}-208$ and free protons at $Q^2 = 10\,\text{GeV}^2$. The results demonstrate that nonlinear effects become increasingly significant with larger $A$. The difference between the parton distribution functions grows progressively with decreasing $x$, reaching maximum at the smallest $x$ values.  In Fig. \ref{Fig2}, the evolution of the gluon distributions for Au-$197$ at $Q^2 =  100\,  \text{GeV}^2$ is shown as a function of $ x$. We have presented a comparison of these results with the nCETQ15  \cite{nCTEQ15} and with the Rausch- Guzey- Klasen (RGK) \cite{Guzey1} model  at the corresponding values of 
$Q^2 =100 \text{GeV}^2$.
 As can be seen in these figures, the our numerical results are comparable with the  nCETQ15  and with the RGK  model. In Fig. \ref{Fig2} (left), the lower panel shows the evolution of the $R_G$ distributions at $Q^2 = 100 \,\mathrm{GeV}^2$. For a fixed value of $Q^2$, $R_G$ ($R_G=\frac{G^A_{\rm GLR-MQ-ZRS}}{G^A_{\rm GLR-MQ}}$) increases as $x$ decreases, indicating that the anti-shadowing effect becomes more pronounced at smaller $x$. As $x$ increases, the contributions from anti-shadowing

\begin{figure}[ht]
	\centering
	\includegraphics[width=17cm]{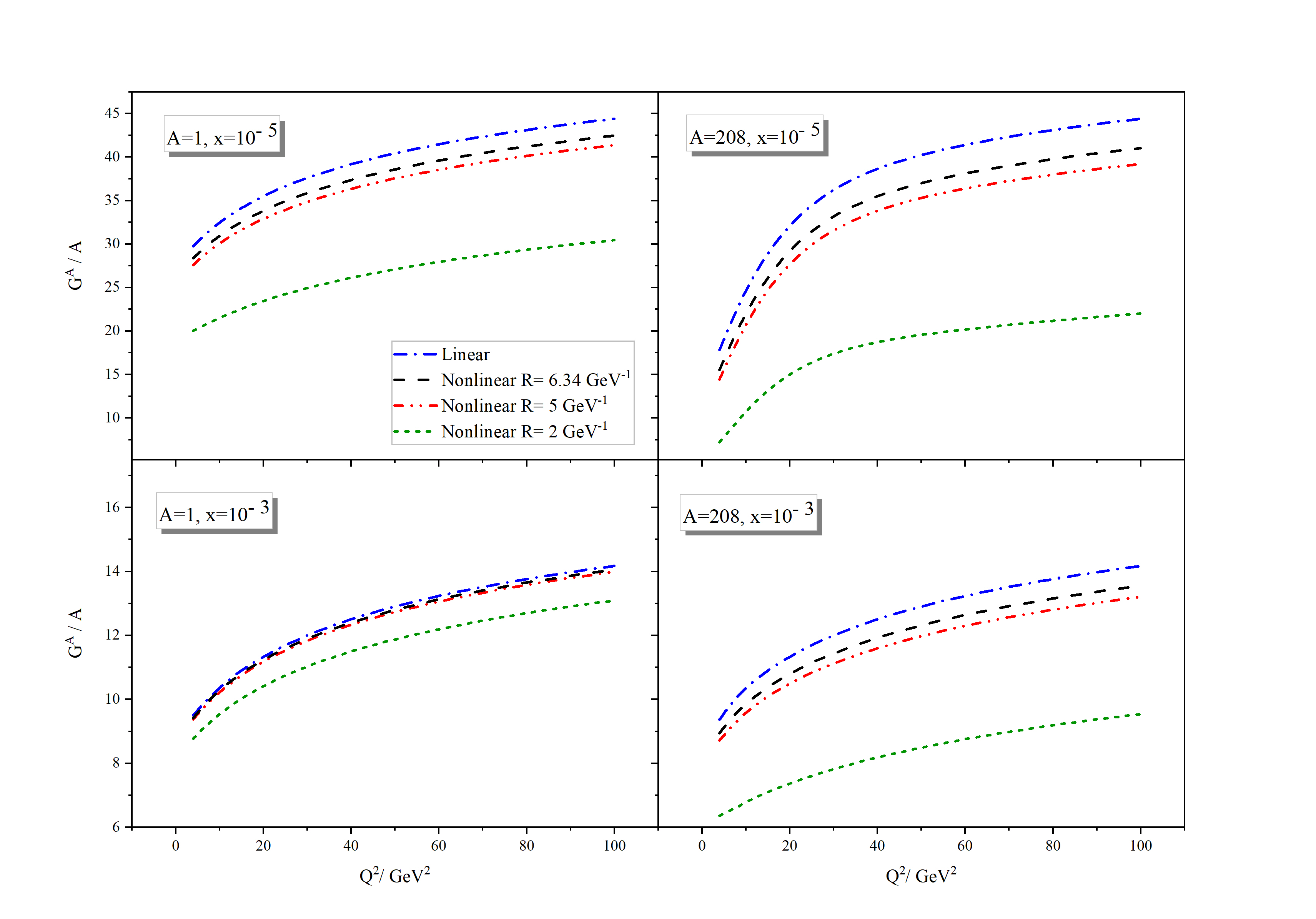}
	\caption{The $Q^2$( ${5}{\text{GeV}^2} \leq Q^2 \leq {100}{\text{GeV}^2}$) evolution of $G(x,Q^2)$  for fixed values of the Bjorken-$x$. The correlation radius $R$ is taken to be $2$, $5$, and $6.34\,  \text{GeV}^{-1}$. The left column shows the gluon distribution function for a nucleon ($A=1$) at $x=10^{-5}$ and $x=10^{-3}$, while the right side shows the corresponding distributions for a heavy nucleus ($A=208$) at the same $x$ values. }
	\label{Fig3}
\end{figure}
\noindent are gradually balanced by the shadowing effect, and $R_G$ approaches unity. In Fig.\ref{Fig3}, we present the $Q^2$ evolution of the gluon distribution function for a nucleon  and a lead nucleus (Pb$-208$) at two Bjorken-$x$ values: $x = 10^{-3}$ and  $x = 10^{-5}$. The results demonstrate that $G(x,Q^2)$ exhibits a significant increase with $Q^2$ for fixed $x$ and $A$. The key observations are:
\begin{itemize}
	\item The gluon distribution function per  nucleon increases as $x$ decreases.
	\item The gluon distribution per nucleon grows at smaller $R$ values.
	\item The gluon distribution per nucleon decreases with increasing mass number $A$.
\end{itemize}

\newpage

\begin{figure}[ht]
	\centering
	\includegraphics[width=17cm]{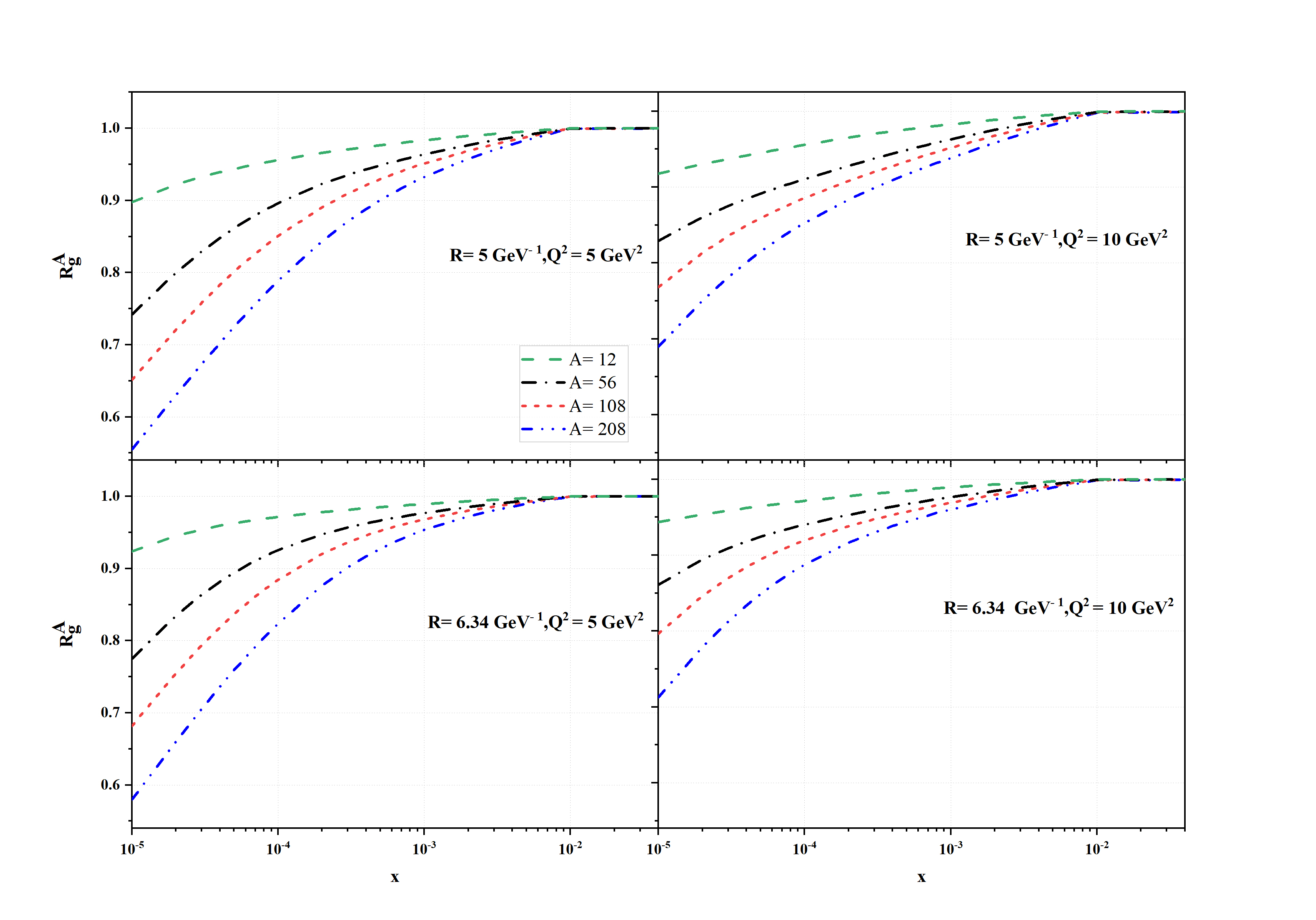}
	\caption{The gluon nuclear modification factors $R_g^A=\frac{ G^A_{NL}}{A G^N_L}$ are calculated at NLO for various nuclei ( $\text{C}-12$ , $\text{Fe}-56$,  $\text{Ag}-108$ and $\text{Pb}-208$) at $Q^2 =5\, \text{GeV}^2$ and\,$Q^2 =10 \, \text{GeV}^2$  with  $R =5\,  \text{GeV}^{-1}$, $R =6.34\, \text{GeV}^{-1}$  }
	\label{Fig4}
\end{figure}
\noindent In Fig. \ref{Fig4}, we present our results for shadowing effects in the nonlinear gluon distribution of nuclei $\frac{1}{A}\frac{G_{A}(x,Q^{2})}{G_{N}(x,Q^{2})}$ at $Q^{2} = 5~\text{GeV}^{2}$ and $Q^{2} = 10~\text{GeV}^{2}$ for light and heavy nuclei including C-12, Fe-56, Ag-108 and Pb-208 at $R = 5 \,  \text{GeV}^{-1}$ and $6.34 \, \text{GeV}^{-1}$. We observe that, as expected,  the shadowing effects are important at lower $Q^{2}$ and smaller $R$ values,  as theoretically expected. The shadowing magnitude decreases with decreasing $x$ and increasing atomic number $A$ ~\cite{Muhammadi}. The obtained results show good agreement with the gluon shadowing corrections reported in Reference \cite{Nemchik}, which account for the $|qqG\rangle$ Fock state contribution in the photon-gluon system. This consistent pattern is also evident across multiple established phenomenological frameworks, including the GBW, KST \cite{Tarasov}, BGBK, and IP-sat models, demonstrating the robustness of the observed behavior. Based on Fig.\ref{Fig4}, it is clear that nonlinear effects become more significant as $A$ increases, particularly at low $x$ and $Q^2$ values. 
\begin{figure}[ht]
	\centering
	\includegraphics[width=12cm]{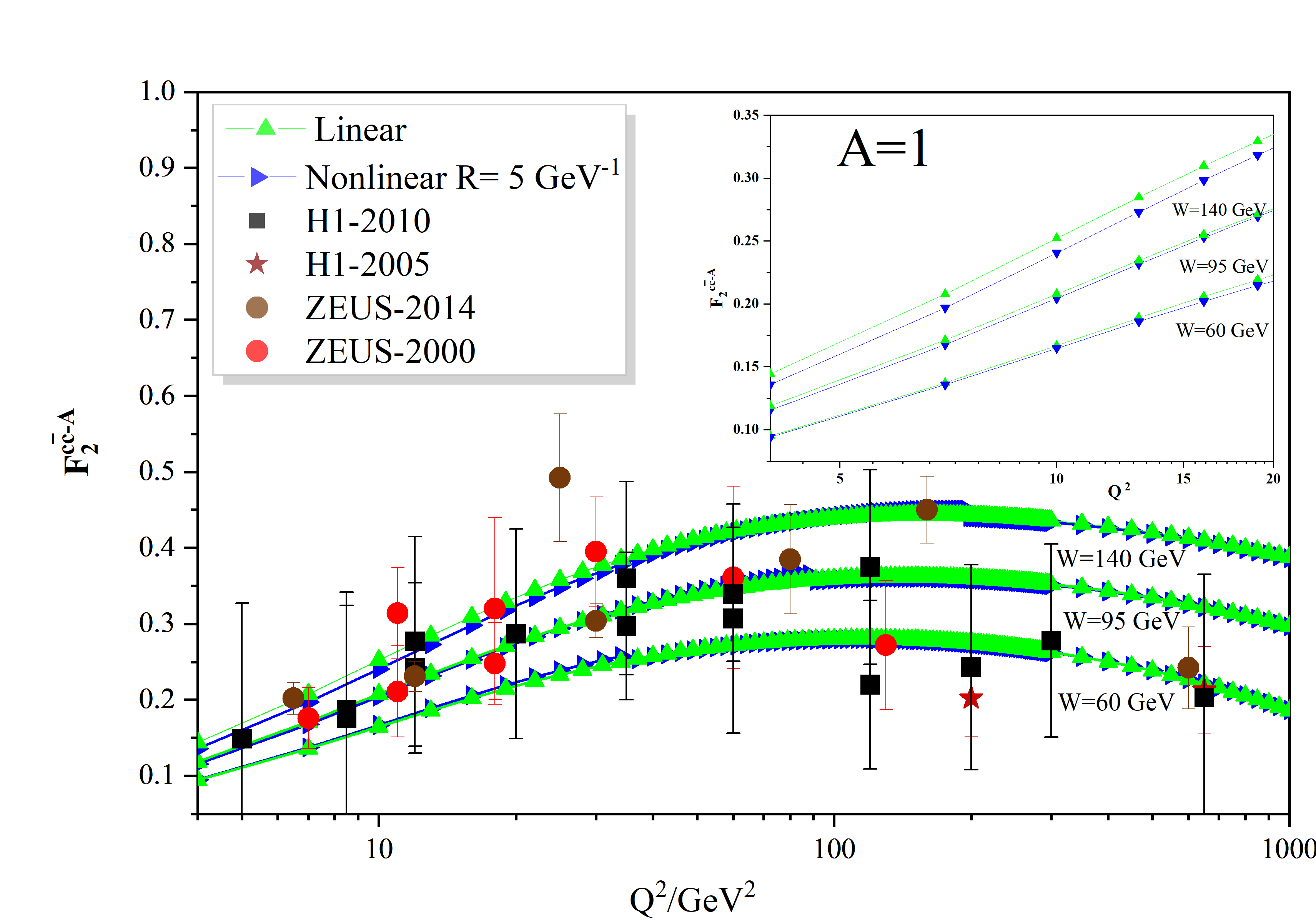}
	\caption{The theoretical prediction for the linear and nonlinear structure function $F^{c\bar{c}}_2 (x, Q^2 )$ at LHeC center-of-mass energy($\sqrt{s} = 1.3\, \text{TeV}$) with $\mu^2 = Q^2 + 4m^2_c$ as a function of variable $Q^2$  for three  different values of $W$, viz. $60$, $95$,  and $140$, respectively.
		Experimental data are taken from the H1 \cite{H1-2005,H1-2010} and ZEUS \cite{ZEUS2014,ZEUS-2001} Collaborations  as accompanied by total errors. }
	\label{Fig5}
\end{figure}
A decrease in this ratio is called shadowing, while an increase is referred to as anti-shadowing. The anti-shadowing effect occurs due to coherent multiple scattering  processes, which enhance 	nuclear effects proportionally to $A^{1/3}$, as shown in references \cite{Boroun1,Boroun2,Boroun3,Boroun4,Boroun5,Boroun6,Boroun7,Boroun8,Boroun9,Boroun10,Boroun11,Boroun12,Boroun13,Boroun14,Boroun15,Guo,Golec}.
In Fig.~\ref{Fig5},  the results for the nonlinear corrections as a function of $Q^2$ on the charm structure function $F_2^{c\bar{c}}$ for $A = 1$ are shown for three different values of $W$, namely $60$, $95$ and  $140  \, \text{GeV}$, respectively, at the NLO approximation with $\mu^2 = Q^2 + 4 m_c^2$ in the kinematic range of the LHeC colliders. As we can see, the nonlinear corrections to the parametrization of structure functions become apparent as $Q^2$ decreases and vanish as $Q^2$ increases. For $W = 60 \, \text{GeV}$ , the corresponding values of $x $ ($x=\frac{Q^2}{W^2}$) are relatively large; therefore, the differences between the linear and nonlinear calculations are small. As $W$ increases to $140  \, \text{GeV}$, the values of $x$ decrease and the linear and nonlinear results start to differ more noticeably. The nonlinear effects are particularly significant for very small $x$ ($x \lesssim 10^{-4}$), and their impact diminishes as $x$ increases. As observed, our numerical results are comparable with the experimental data from HERA  \cite{H1-2010,H1-2005,ZEUS-2001,ZEUS2014}.

\newpage
\begin{figure}[ht]
	
	\centering
	\includegraphics[width=17cm]{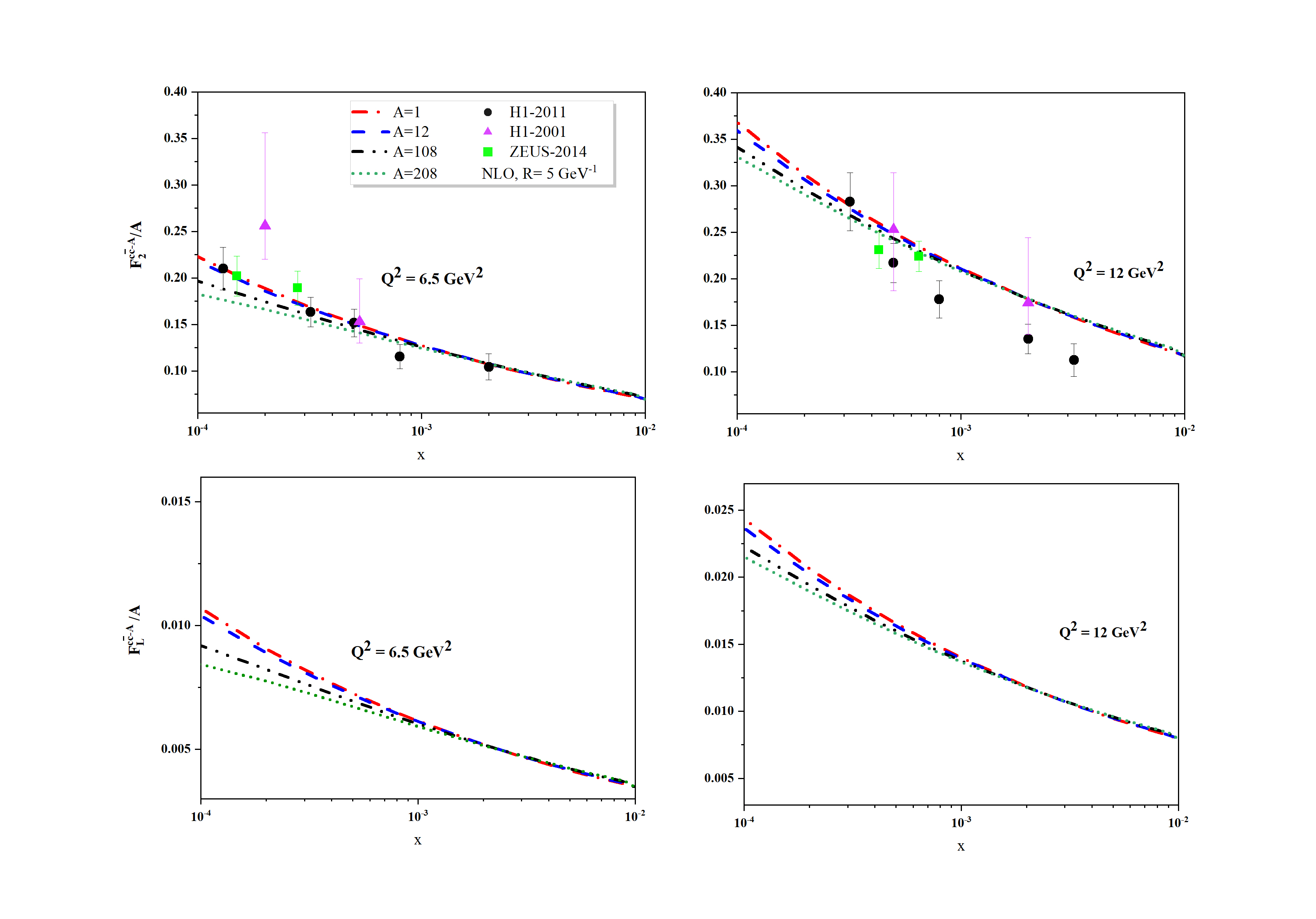}
	\caption{Results of the charm structure function per nucleon $F^{c\bar{c}-A}_2 /A$ and $F^{c\bar{c}-A}_L /A$ as a function of $x$ for C-$12$, Ag-$108$, Pb-$208$ nuclei and the free proton at $Q^2=6.5$ and $12\, \text{GeV}^2$ calculated in our analysis at the renormalization scale $\mu ^2 = Q^2 + 4m^2_c$ at $R = 5 \, \text{GeV}^{-1}$ in the NLO approximation. The results are compared with the H1 \cite{H1-2001,H1-2011} and ZEUS \cite{ZEUS2014} Collaborations, accompanied by total errors.} 
	\label{Fig6}
\end{figure}
\begin{figure}[ht]
	\centering
	\includegraphics[width=15cm]{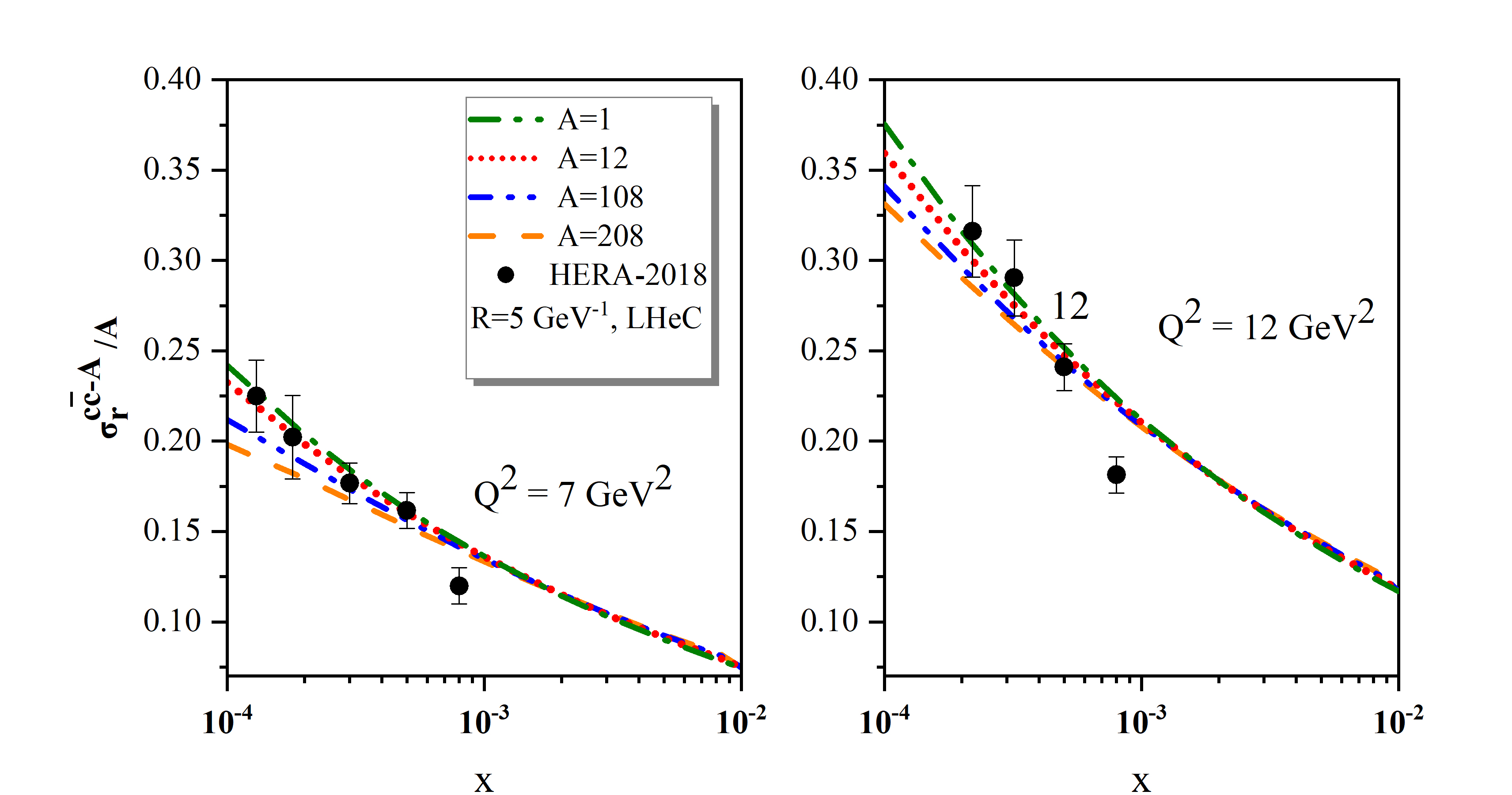}
	\caption{ The nonlinear behavior of the  charm  reduced cross section  ($\sigma_r^{c\bar{c}-A}/A$) are shown as a function of the momentum fraction $x$ at $Q^2 = 7$ and $12\, \text{GeV}^2$ for  C-$12$ , Ag-$108$, Pb-$208$ nuclei and free proton at the NLO approximation in the LHeC  COM energies ,  with $\mu^2 = Q^2 + 4 m^2_c$ for $0 < y < 0.95$. These results are obtained at $R  = 5 \,\text{GeV}^{-1} $. The nonlinear corrections  per nucleon have been compared with experimental data from the HERA Collaborations\cite{H1-2018}, including total errors.  }
	\label{Fig7}
\end{figure}
\begin{figure}[ht]
	\centering
	\includegraphics[width=15cm]{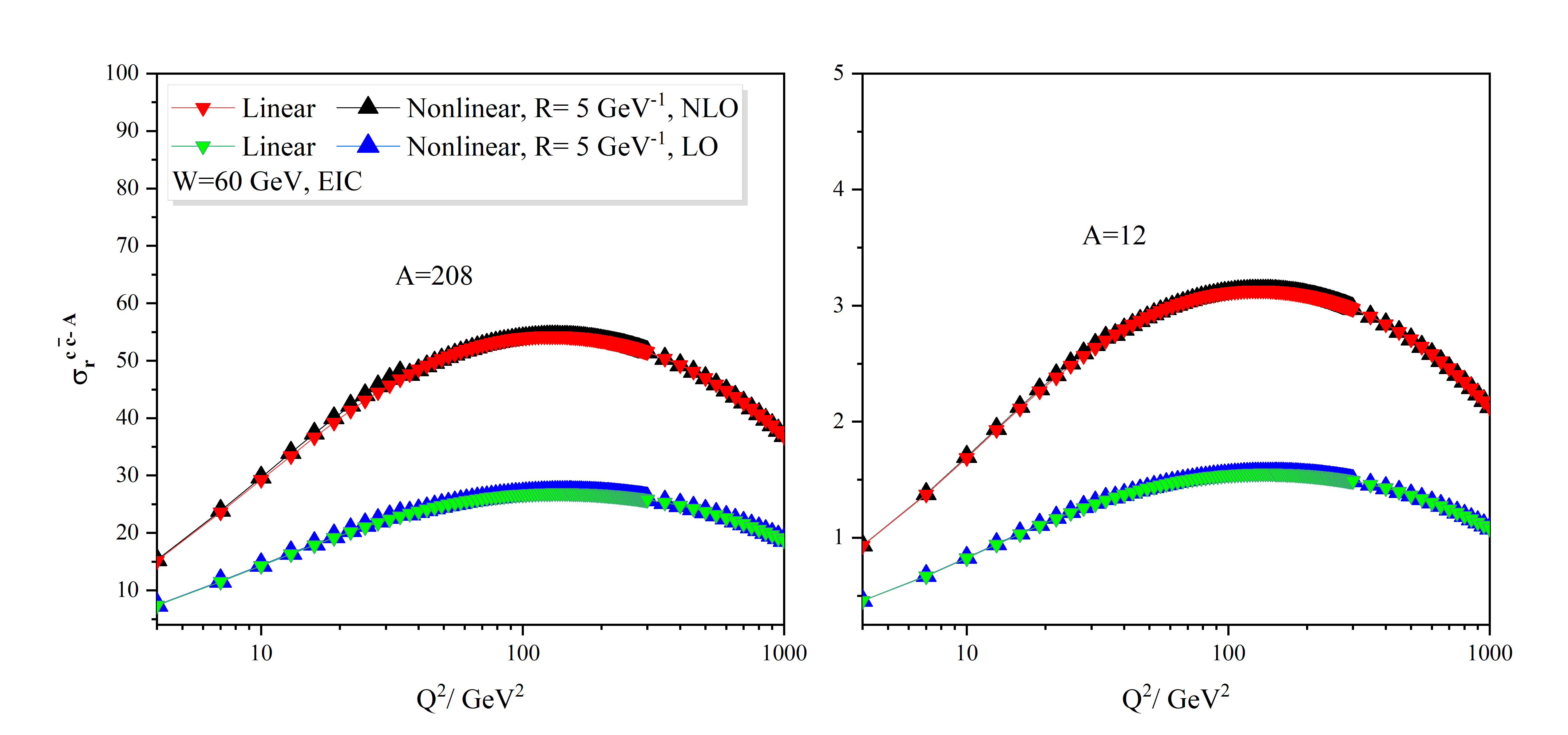}
	\caption{Our results of the  linear and nonlinear charm reduced cross section  ($\sigma_r^{c\bar{c}}$)  at $\sqrt{s} =  89.4\, \text{GeV}$ (EIC center-of-mass energy)   are presented for C-$12$ and Pb-$208$ nuclei. The results are taken at fixed  $W$ and plotted as a function of $Q^2$.  }
	\label{Fig8}
\end{figure}

\noindent Fig. \ref{Fig6} illustrate  the $F^{c\bar{c}-A}_2 /A$  and  $F^{c\bar{c}-A}_L /A$   charm structure functions of nuclei per nucleon as a function of $x$  for C-$12$, Ag-$108$, Pb-$208$ and free proton at $R=5 \, \text{GeV}^{-1}$, for two different values of $Q^2$ viz. $6.5$ and $12\, \text{GeV}^2$ respectively.  We observe that the results of  $F^{c\bar{c}-A}_2 /A$\,( $F^{c\bar{c}-A}_L /A$) for light and heavy nuclei, at very low $x$ increase. Also, the charm structure functions of nuclei per nucleon increases with increasing $Q^2$.
As  is observed, the our numerical results are comparable with experimental data of HERA \cite{H1-2001,H1-2011,ZEUS2014} predictions.\\
In Fig. \ref{Fig7}, we check the sensitivity of $A$ and $Q^2$ on our results. The nonlinear behavior of the  charm  reduced cross section ($\sigma^{c \bar{c}-A}_r/A$)   for   $R = 5 \,\text{GeV}^{-1}$ are shown as a function of the momentum fraction $x$ at $Q^2 = 7$ and $12 \, \text{GeV}^2$ for C-$12$ , Ag-$108$, Pb-$208$ and free proton at
\\
 \noindent the NLO approximation in the LHeC  ($\sqrt{s} = 1.3\,  \text{TeV}$) colliders,  with $\mu^2 = Q^2 + 4 m^2_c$ for $0 < y < 0.95$. We observe a  rise of $\sigma^{c \bar{c}-A}_r/A$ towards small-x as we decrease the value of $ A$. The numerical predictions for nonlinear cross sections in the kinematics regime of very small- $x$  and for inelasticity  $0 < y < 0.95$ at the  NLO approximations with  $\mu^2 = Q^2 + 4 m^2_c$ at the LHeC Colliders are compared with HERA Collaborations data \cite{H1-2018}. The effects of the nonlinear corrections  are  particularly pronounced at low values of $R$. Our findings for the reduced cross section demonstrate a good agreement with the experimental data at the chosen renormalization scale.\\
In Fig. \ref{Fig8},  the values of $\sigma_r^{c\bar{c}-A}$ are shown for $\text{C}-12$ and $\text{Pb}-208$ at LO and NLO approximations to facilitate comparison. As observed, for heavy nuclei at $W = 60\,\text{GeV}$, the corresponding values of $x$ are relatively large, so the differences between linear and nonlinear calculations are small.  This behavior highlights that nonlinear effects, such as shadowing and antishadowing, are more pronounced at small $x$ and for heavier nuclei, while they are negligible at larger $x$ 
 
\noindent or lighter nuclei. \\

 In conclusion,  this  study examines the behavior of gluon distributions
using the  GLR-MQ-ZRS  nonlinear equation for light and heavy
nuclei at small-$ x$. The results show sensitivity to the atomic number $A$ and  the correlation radius  ($R$) of interacting gluons. As $x$  decreases and  $A$ increases, the impact of gluon recombination becomes more pronounced. Additionally, the analysis suggestes that the suppression of the gluon distribution function is more significant at small-$x$ and low $Q^2$, compared to small-$ x$  but higher $ Q^2$. These findings suggest that inclusive observables are affected by nonlinear effects.\\ The nonlinear corrections to the nuclear gluon distribution are comparable to the evolution
method from the RGK model \cite{Guzey1}. In the following, we explored the phenomenological predictions of  heavy quark pair production at center-of-mass energies $\sqrt{s} = 1.3\, \text{TeV}  $ and  $\sqrt{s} = 89.4\, \text{GeV}  $ corresponding to the LHeC and EIC colliders in the collinear generalized double asymptotic (DAS) approach \cite{DAS}. We studied the evolutions of  nonlinear heavy quark pair production cross section, structure functions $F_2^{c \bar{c}-A} $ and $F_L^{c \bar{c}-A} $ for nuclei  with respect to the variables $x$ and $Q^2$  at the NLO approximation.\\ 
We observed that the nonlinear corrections are noticeable in the small-$x$ kinematic region, especially at low $Q^2$ values and small nuclear radii ($R$). The results computed in this study show compatibility with results from various collaboration groups such as nCTEQ15 \cite{nCTEQ15}, as well as experimental data from the ZEUS \cite{ZEUS-2001,ZEUS2014} and H1 \cite{H1-2001,H1-2005,H1-2010,H1-2011,H1-2018} collaborations.\\

\section{Appendix}
In NLO, we have used the compact form of the co-efficient functions
in high energy regime ($\zeta << 1$) according to \cite{bkg}
\begin{equation}
	C_{k,g}^{(1)}(x,\xi)=\beta[R_{k,g}^{(1)}(1,\xi)+4C_{A} C_{k,g}^{(0)}(1,\xi)L_{\mu}], \quad L_{\mu}=\ln(\frac{4m^{2}}{\mu^{2}})
\end{equation}

with
\begin{align}\label{2}
	R_{2,g}^{(1)}(1,\xi) &=\frac{8}{9}C_{A}[5+(13-10\xi)J(\xi)+6(1-\xi)I(\xi)],
	\nonumber \\ R_{L,g}^{(1)}(1,\xi) &=-\frac{16}{9}C_{A}x_{1}[1-12\xi-(3+4\xi(1-6\xi))J(\xi)+12\xi(1+3\xi)I(\xi)],
	\nonumber \\ C_{2,g}^{(0)}(1,\xi) &=\frac{2}{3}[1+2(1-\xi)J(\xi)],
	\nonumber \\ C_{L,g}^{(0)}(1,\xi) &=\frac{4}{3}x_{1}[1+6\xi-4\xi(1+3\xi)J(\xi)],
\end{align}
where
\begin{align}\label{3}
	I(\xi) &=-\sqrt{x_{1}}[\zeta_{2}+\frac{1}{2}\ln^{2}(t)-\ln(\xi x_{1}) \ln(t)+2Li_{2}(-t)],
	\nonumber \\ J(\xi) &=-\sqrt{x_{1}}\ln(t), \quad t= \frac{1-\sqrt{x_{1}}}{1+\sqrt{x_{1}}}, \quad\beta^{2} =1-\frac{4\xi x}{1-x},
	\nonumber \\ L(\beta)&=\frac{1}{\beta}\ln\frac{1+\beta}{1-\beta}
	\nonumber \\ Li_{2}(x) &=-\int_{0}^{1}\frac{dy}{y}\ln(1-xy),
\end{align}
where $Li_{2}(x)$ is the dilogarithmic function.
\section*{ACKNOWLEDGMENTS}

Authors are grateful to  G.R. Boroun  for useful discussions and comments on the paper.

\end{document}